\documentclass[aps,preprint,showkeys,
superscriptaddress,
preprint,
amsmath,amssymb,
aps,
]{revtex4-1}

\usepackage[latin1]{inputenc}
\usepackage{graphicx}% Include figure files
\usepackage{dcolumn}% Align table columns on decimal point
\usepackage{bm}% bold math
\usepackage{color}

\begin{document}

\date{\today}

%\preprint{APS/123-QED}

\title{Partial control of delay-coordinate maps }% Force line breaks with \\

\author{Rub\'en Cape\'ans}
\email{ruben.capeans@urjc.es}
\affiliation{Departamento de F\'isica, Universidad Rey Juan Carlos, Tulip\'an s/n, 28933 M\'ostoles, Madrid, Spain}

\author{Juan Sabuco}
\email{juan.sabuco@urjc.es}
\affiliation{Departamento de F\'isica, Universidad Rey Juan Carlos, Tulip\'an s/n, 28933 M\'ostoles, Madrid, Spain}

\author{Miguel A. F. Sanju\'an}
\email{miguel.sanjuan@urjc.es}
\affiliation{Departamento de F\'isica, Universidad Rey Juan Carlos, Tulip\'an s/n, 28933 M\'ostoles, Madrid, Spain}
\affiliation{Institute for Physical Science and Technology, University of Maryland, College Park, Maryland 20742, USA}

\begin{abstract}

Delay-coordinate maps have been widely used recently to study nonlinear dynamical systems, where there is only access to the time series of one of their variables. Here, we show how the partial control method can be applied in this kind of framework in order to prevent undesirable situations for the system or even to reduce the variability of the observable time series associated with it. The main advantage of this control method, is that it allows to control delay-coordinate maps even if the control applied is smaller than the external disturbances present in the system. To illustrate how it works, we have applied it to three well-known models in Nonlinear Dynamics with different delays such as the two-dimensional cubic map, the standard map and the three-dimensional hyperchaotic H\'{e}non  map. For the first time we show here how hyperchaotic systems can be partially controlled.

\end{abstract}

\maketitle

\section{Introduction}

In Dynamics we usually study physical systems whose present state is completely defined by a set of $m$ variables $x_1, x_2, x_3,..., x_m$, where the law governing its evolution $f(x_1, x_2, x_3,..., x_m)$ is known as well. The set of all the possible values that these variables can take constitute the phase space of the system, which is usually represented geometrically as a Cartesian coordinate system. In this way every possible state can be identified with a unique point in the phase space. However, there are several situations where this approach is not possible. We find one example of these situations in systems with memory \cite{DDE}, where the future state of the system does not depend only on the present state, but also on other previous states. This kind of dynamical systems are usually referred to as systems with delay in the literature. In other physical systems, however, it is impossible to have access to the information of all the variables that constitute its phase space. In this kind of situation the delay reconstruction method \cite{Geometry_time_series, Takens_time_delay} is used to study the dynamics of the underlying system. To do so, a time series of scalar measurements of one of the variables is taken in order to reconstruct the dynamics of the phase space and estimate its delay-coordinate map \cite{Embedology,Embedology2}. This method has been extensively applied to study nonlinear time series.

In systems with delay, it is common to find dynamics with chaotic behavior. This kind of dynamics allows the system to visit many different states in the phase space. But in some situations some of these states might be dangerous for the future evolution of the system or might just enhance the appearance of large extremes in the time series. In the former case, the system behaves chaotically for a while, but eventually escapes to an external attractor that might be harmful for the system (like for example an extinction \cite{Ecology}). In the latter, the presence of the extremes introduces a high volatility in the time series that can have a negative effect in some situations \cite{Das}. In order to avoid this kind of behaviors, some control has to be applied. Unlike classical control methods that pursue to transform chaotic trajectories into periodic ones \cite{OGY}, in these other cases the main goal is to avoid unwanted behaviors in the presence of a chaotic dynamics. To achieve this goal, it has been proposed recently the partial control method \cite{Automatic,Asymptotic}, which is applied in chaotic systems with the aim of suppressing some undesirable states of the system in the presence of any kind of disturbances. Due to the disturbances, and the tiny control used, it is not possible to guide the trajectory to an specific target. We only can keep the trajectory in some region of the phase space, and that is why it is called ``partial control''. This control method is minimally invasive and has been tested in many different scenarios. For example, it was shown in \cite{Das} the possibility of suppressing extremes of a variable in an economic model, reducing the volatility. It has also been shown in several examples the possibility of avoiding escapes in situations where the chaotic behavior is transient \cite{Ecology,Cancer,Lorenz}.

\begin{figure}
\includegraphics [trim=0cm 0cm 0cm 0cm, clip=true, width=0.95\textwidth]{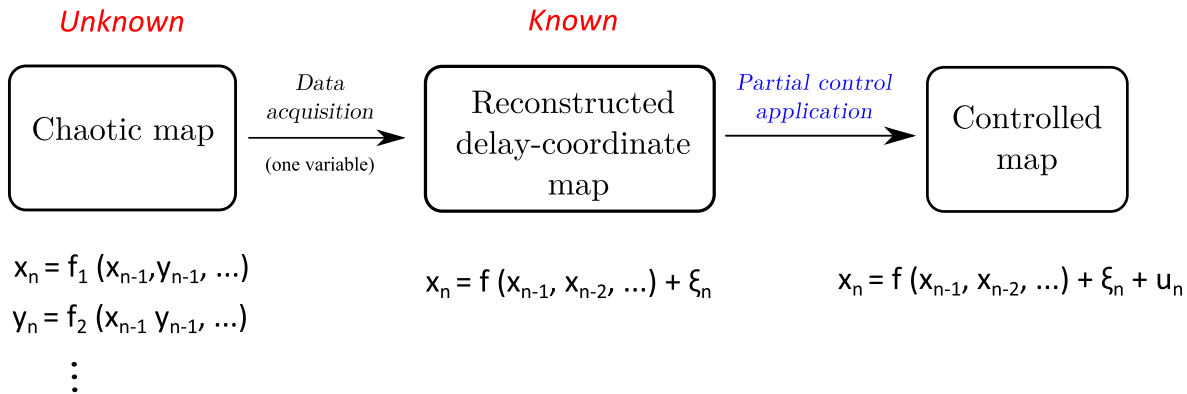}\\
\centering
\caption{\textbf{Conceptual framework.} From left to right. Step $1$: data acquisition from a chaotic system. We assume here that only one variable is observable. Step $2$: using embedding and parametric reconstruction techniques, construct a delay-coordinate map. The term $\xi_n$ represents a disturbance term that encloses all possible deviations from the real dynamics. Step $3$: apply the partial control method introducing and additive control term $u_n$ acting on the observable variable. In this work we assume that we already possess the knowledge of the delay-coordinate map.}
\label{-1}
\end{figure}

The main goal of this paper is to show how the partial control method can be applied to delay-coordinate maps under very mild assumptions. We consider here a delay-coordinate map under external additive disturbances $f(x_n, x_{n-1}, ...)+\xi_n$, where the control can also be applied in an additive way $f(x_n, x_{n-1}, ...)+\xi_n+u_n$. This kind of framework is the one that is usually found after using the delay reconstruction method to study the phase space dynamics of a chaotic system, but can also be found in others systems with delay. This method is normally employed in order to make short term predictions of the underlying time series by estimating the associated delay-coordinate map, which can be done by fitting it to some polynomial \cite{Predicting} or using other alternative equation-free approaches like empirical dynamic modeling  \cite{EDM}. However, the delay-coordinate maps can also be employed to control chaotic time series. This makes us believe that the partial control method might be used in all these situations in order to avoid certain undesired behaviors.

The structure of the paper is as follows. In Section $2$, we describe how the partial control method is applied to delay-coordinate maps. In Section $3$ we present several examples of how to apply this control method to three paradigmatic delay-coordinate maps found in the literature, including one hamiltonian map and one hyperchaotic system. Finally, some conclusions are drawn.

\section{Application of the partial control method to Delay-coordinate maps}

The partial control method has been successfully applied in several situations \cite{Automatic,Asymptotic,Ecology,Cancer,Lorenz}, where the control was applied on the variables of the system, and also where the control was applied on some parameter of the system \cite{Parametric}. The method was originally defined on maps, however it also applies well in the case of flows \cite{Lorenz}, taking a suitable discretization of the dynamics.

In this work we pursue the application of the method to delay-coordinate maps. These maps are usually expressed in the following way:
\begin{equation}
\begin{array}{l}
x_{n}=f(x_{n-1},x_{n-2}\ldots x_{n-m}).
\end{array}
\end{equation}

We consider here the problem of controlling this kind of maps possessing a chaotic behaviour.  It may seem that the implementation of the partial control method on time delay systems do not differ much from the previous works \cite{Automatic,Asymptotic,Ecology,Cancer,Lorenz}, however there is a critical difference since the control can only be applied in the present state $x_n$, (is not physically possible to control the past states $(x_{n-1},x_{n-2}\ldots)$), and therefore the control of the system involves the introduction of a new approach.

Following the scheme of the partial control method we consider that the system can be modelled as:
\begin{equation}
\begin{array}{l}
 x_{n}=f(x_{n-1},x_{n-2}\ldots x_{n-m})+\xi_n +u_n,
\end{array}
\end{equation}
where $\xi_n$ is the disturbance affecting the state $x_{n}$, and $u_n$ is the respective control applied. These values are constrained
\begin{equation}
\begin{array}{l}
|\xi_n|\leq\xi_0, ~~~~~~|u_n|\leq u_0.\nonumber
\end{array}
\end{equation}

We will show here that it is possible to control the systems with control values $0<u_0<\xi_0$, which is one of the most remarkable results of this method.

We recall here that in the experimental case, this delay-coordinate map would be obtained from the collected data, after the application of some embedding or model reconstruction techniques. However this task is not the main goal of this work. What we show here is the application of the partial control method to this kind of maps, so we assume here that the delay-coordinate map is already known. To illustrate this point, a small scheme (Fig.~\ref{-1}) was added.

\begin{figure}
\includegraphics [trim=0cm 0cm 0cm 0cm, clip=true, width=0.8\textwidth]{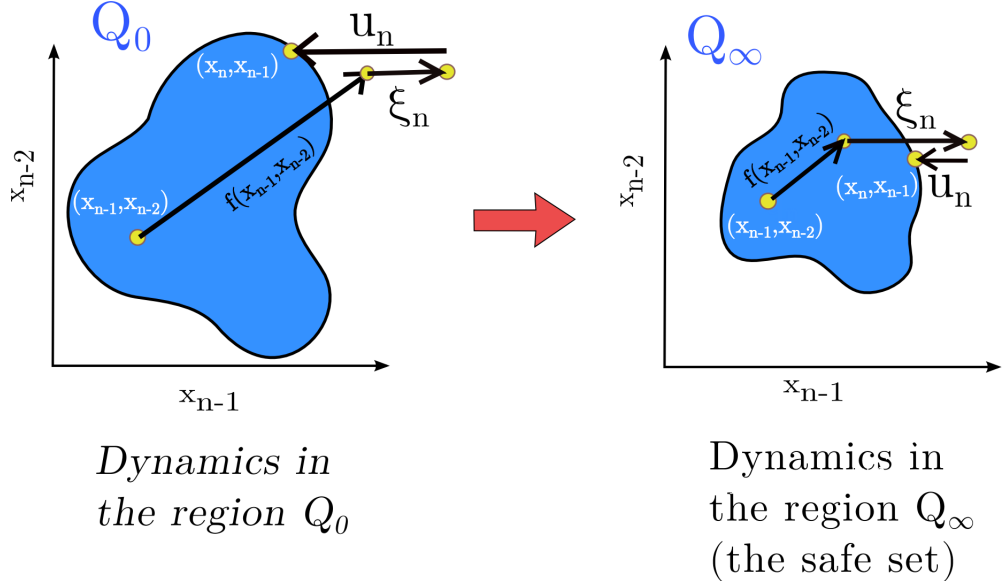}\\
\centering
\caption{\textbf{Dynamics in $Q_0$ and $Q_\infty$.} The left side shows an example of a $2D$ region $Q_0$ (in blue) in which we want to keep the dynamics described by $x_{n}=f(x_{n-1},x_{n-2})+\xi_n+u_n$. We say that $|\xi_n| \leq \xi_0$ is a bounded disturbance affecting the map, and $u_n$ is the control chosen so that $q_{n+1}$ is again in $Q_0$. Notice that disturbance and control arrows are drawn parallel to current state of the variable since only the present state is affected by them. To apply the control, the controller only needs to measure the state of the disturbed system, that is $[f(x_{n-1},x_{n-2})+\xi_n]$. The knowledge of $f(x_{n-1},x_{n-2})$ or $\xi_n$ individually is not required. The right side of the figure, shows the region $Q_\infty \subset Q_0$ (in blue), called the \emph{safe set}, where each $(x_{n-1},x_{n-2}) \in Q_\infty$ has $(x_{n},x_{n-1}) \in Q_\infty$ for some control $|u_n|\leq u_0$, which is chosen depending on $\xi_n$. Notice that the removed region does not satisfy $|u_n|\leq u_0$.}
\label{0}
\end{figure}

Once we know the delay-coordinate-map,  all we have to do to apply the partial control method is to define the region $Q_0$ of the phase space $(x_{n-1},x_{n-2}\ldots)$ where we want to keep the trajectories. Then we have to fix the upper value of the disturbance $\xi_0$, and the upper value of the control $u_0$. Next the safe set is computed. This set is formed by the points $(x_{n-1},x_{n-2}\ldots)$ belonging to the region $Q_0$,  where the image $x_{n}=f(x_{n-1},x_{n-2}\ldots)+\xi_n+u_n$ can be put back again on the safe set by using a control $|u_n|\leq u_0$. In Fig.~\ref{0} we illustrate the controlled dynamics in the region $Q_0$ and the safe set $Q_\infty$.  Notice that, due to the fact that the control and disturbance affects the present state of the variable, then they are applied in the current axis direction.
%In the case of delayed coordinate maps, the phase space is represented by the present state and theirs delayed states for example $(x_{n-1},x_{n-2},x_{n-3})$. Taking into account that only the present state is physically accessible to the control, and that we are considering a random disturbance affecting this state, the fatten and shrinking process of the Sculpting Algorithm will be made only in this dimension of the phase space as it is shown in Fig.~\ref{0}.

To compute the safe set, it has been developed a recursive algorithm called the \emph{Sculpting Algorithm}~\cite{Automatic}, which evaluates the points from $Q_0$ and remove them if they do not satisfy the safe condition. We have modified it here to apply it to delay-coordinate maps. The new algorithm converges, after some steps, when all the points remaining are safe considering that we only apply control in the present state of the variable. The $ith$ step of this algorithm can be summarized as follows:

\begin{enumerate}
  \item  Morphological dilation of the set $Q_i$ by $u_0$ along the $x_{n-1}$ direction, obtaining the set denoted by $Q_i+u_0$.
  \item  Morphological erosion of set $Q_i+u_0$ by $\xi_0$ along the $x_{n-1}$ direction, obtaining the set denoted by $Q_i+u_0-\xi_0$.
  \item  Let $Q_{i+1}$ be the points $(x_{n-1},x_{n-2}\ldots)$ of $Q_i$, so that $f(x_{n-1},x_{n-2}\ldots)$ maps inside the set $Q_i+u_0-\xi_0$.
  \item  Return to step $1$, unless $Q_{i+1}=Q_i$, in which case we set $Q_\infty=Q_i$. We call this final region, the \textit{safe set}. Note that if the chosen $u_0$ is too small, then $Q_\infty$ may be the empty set, so that a bigger value of $u_0$ must be chosen.
\end{enumerate}

Since  computational resources are limited,  only a finite amount of points in $Q_0$ can be evaluated. In this work we have used a rectangular or a cubic grid covering $Q$, but other choices are possible. Certainly higher resolutions give a more accurate safe set. However, we have found that beyond a critical resolution of the grid of $Q$ and $\xi$, the safe set remains the same. For that reason, we recommend to compute the safe set with increasing resolutions until the safe set remains practically unchanged.

In order to show that the method can be applied on different chaotic maps, we have chosen three examples of well-known chaotic maps to illustrate it. We do not reproduce here the embedding and reconstruction model step, since is not the goal of this paper. Instead of that, we have deduced by simple calculation, the expression of the delay-coordinate maps. Next, we apply the control scheme with the aim of keeping the orbits in a desirable region of the phase space.

\subsection{The two-dimensional cubic map}

We consider here the system  given by:
\begin{equation}
\begin{array}{l}
x_{n}=y_{n-1}\\
y_{n}=-bx_{n-1}+ay_{n-1}-y_{n-1}^3,
\end{array}
\label{eq1}
\end{equation}
which represents the two-dimensional cubic map \cite{Holmesa}.

\begin{figure}
\includegraphics [width=1\textwidth]{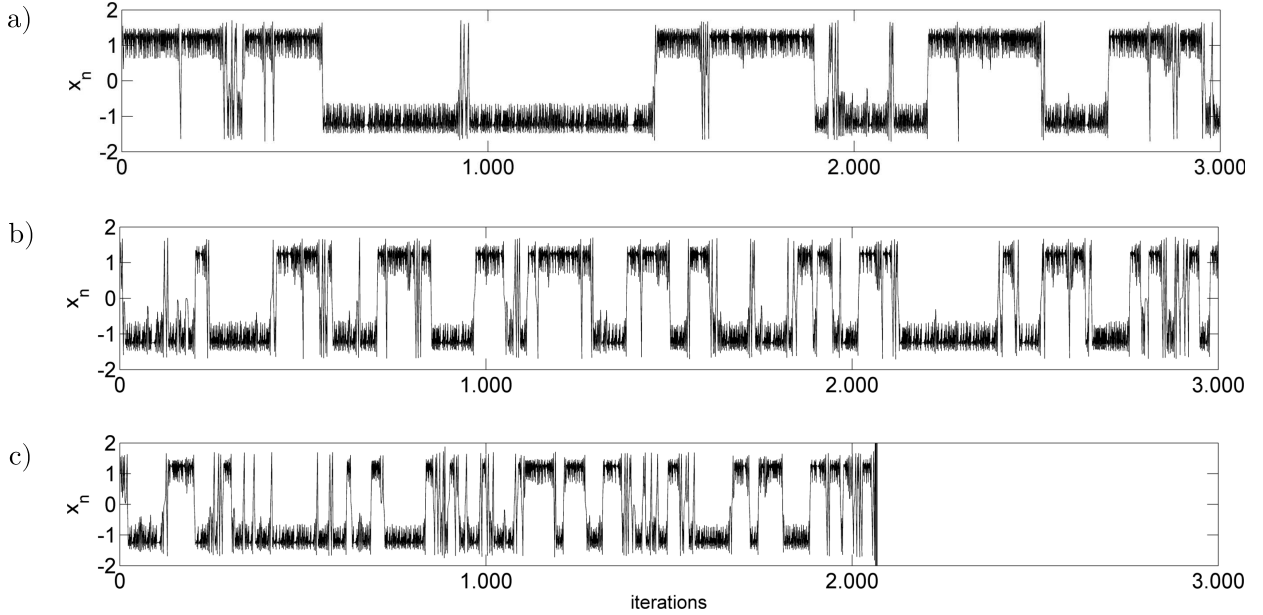}\\
\centering
\caption{\textbf{Time series of the two-dimensional cubic map for different disturbances.} a) Time series of variable $x_n$ with no disturbance affecting it. b) Time series with $|\xi_n|\leq\xi_0=0.02$ affecting the map. c) Time series with $|\xi_n|\leq\xi_0=0.20$ affecting the map. After some iterations the trajectory escapes towards $-\infty$.}
\label{4}
\end{figure}

This two-dimensional cubic map depends on two parameters and exhibits chaos for different values of them. We have selected here the values $a =2.75$ and  $b=0.2$. For this choice of parameters, we have represented in Fig.~\ref{4}a  an example of the time series of the variable $x_n$ without the influence of noise. Here, we can see that the trajectories oscillate between two well differentiated regions (top and bottom), where the transitions between them occurs after some typical time. However, when we introduce additive disturbances, the frequency of the transitions increases (Fig.~\ref{4}b). And for large disturbances the trajectory eventually escapes toward an external attractor due to the extra energy applied (Fig.~\ref{4}c).

\begin{figure}
\includegraphics [trim=0cm 0cm 0cm 0cm, clip=true,width=0.9\textwidth]{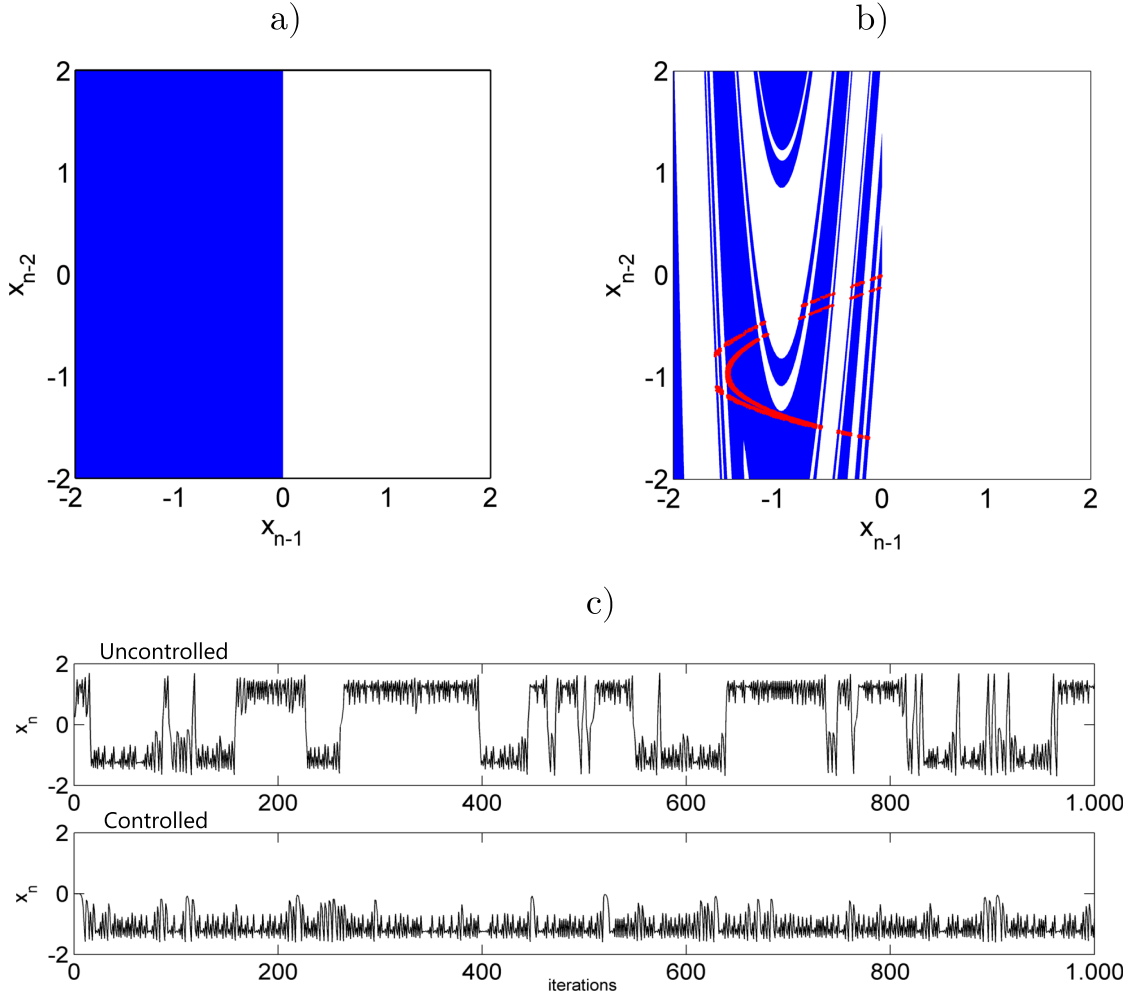}\\
\centering
\caption{\textbf{Safe set and controlled dynamics in the two-dimensional delayed cubic map ($\mathbf{x_{n}=ax_{n-1}-bx_{n-2}-x_{n-1}^3}$).} a)In blue the initial region $Q_0$ where we want to keep the trajectories. b) The safe set obtained with the values of disturbance $\xi_0=0.020$ and control $u_0=0.015$. A grid of $1000 \times 1000$ points has been used. The red dots represent $1000$ iterations of a partially controlled trajectory. c) In the top it is represented an uncontrolled time series affected with $\xi_n\leq\xi_0=0.020$. In the bottom the controlled time series corresponding to the red dots shown in case b.}
\label{5}
\end{figure}

As an example, we assume now that due to experimental restrictions we only see the dynamics of the variable $x_n$ and, with that information, we are interested in keeping the trajectory in the bottom region ($-2<x_n<0$) forever, even in presence of large disturbances.

The form of the reconstructed delay-coordinate map can be deduced by substituting  $y_{n-1}=-bx_{n-2}+ay_{n-2}-y_{n-2}^3$ into Eq. \ref{eq1} and taking into account that $x_{n-1}=y_{n-2}$.

\begin{equation}
\begin{array}{l}

x_{n}= ax_{n-1}-bx_{n-2}-x_{n-1}^3.

\end{array}
\end{equation}

We call this map the \emph{two-dimensional delayed cubic map}. In addition, we add to the model a disturbance term $\xi_{n}$ in order to consider the potential noise present in the data acquisition or also mismatches in the reconstruction model technique.

\begin{figure}
\includegraphics [trim=0cm 0cm 0cm 0cm, clip=true,width=0.9\textwidth]{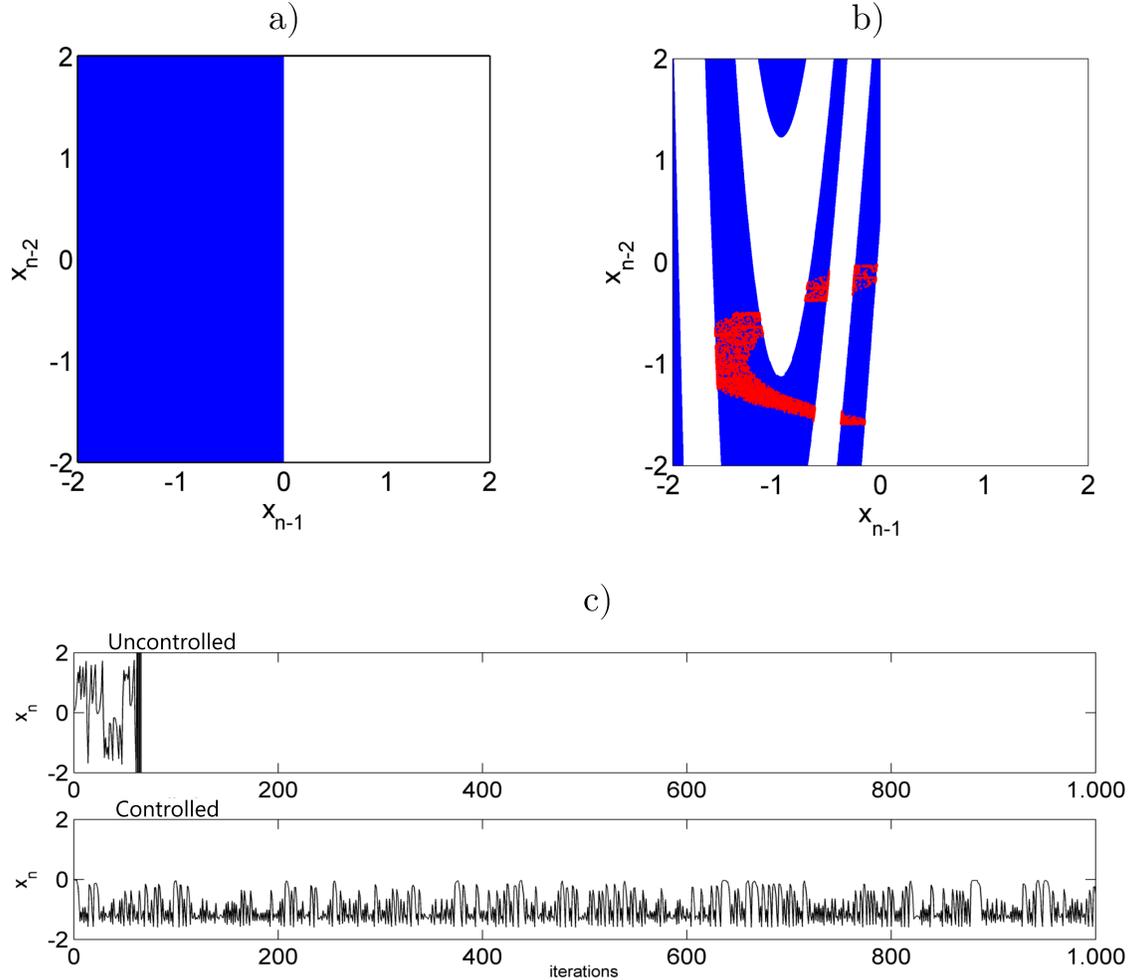}\\
\centering
\caption{\textbf{Safe set and controlled trajectory in two-dimensional delayed cubic map ($\mathbf{x_{n}=ax_{n-1}-bx_{n-2}-x_{n-1}^3}$).} a) In blue the initial region $Q_0$ where we want to keep the trajectories. b) The safe set obtained with the values of disturbance $\xi_0=0.20$ and control $u_0=0.15$. A grid of $1000 \times 1000$ points has been used. The red dots represent 1000 iterations of a partially controlled trajectory. c) In the top it is represented an uncontrolled time series affected with $\xi_n\leq\xi_0=0.20$. In the bottom the controlled time series corresponding to the red dots shown in b)}
\label{6}
\end{figure}

Taking into account the disturbance and the control term $u_n$ in the system, the controlled scheme is given by:
\begin{equation}
\begin{array}{l}
x_{n}=ax_{n-1}-bx_{n-2}-x_{n-1}^3+\xi_{n}+u_n,
\end{array}
\end{equation}
with $|\xi_n|\leq\xi_0$ and $|u_n|\leq u_0$.

To apply the partial control method, the first thing that we have to do is to define the initial $Q_0$ region (Fig.~\ref{5}a) where we want to keep the trajectory. Notice that it is enough to take ($-2<x_{n-1}<0$) to ensure that all successive $x_n$ values remain in this interval.  Then, we select two different  values of $\xi_0$ in order to show how the safe set changes.  The first safe set (Fig.~\ref{5}b) was computed  with the values $\xi_0=0.020$ and  $u_0=0.015$. In the second one (Fig.~\ref{6}b) the values $\xi_0=0.20$ and $u_0=0.15$  were used. In both situations, we have used approximately the smallest possible value of $u_0$. For smaller values, no safe sets exist.

How to use the safe set?. By definition, we start with a point $(x_{n-1},x_{n-2})$ and apply the map $f(x_{n-1},x_{n-2})+\xi_n$. If the image falls inside the safe set no control is applied, if it falls outside, a control is applied to the closest point belonging to the safe set. In Fig.~\ref{5}b and Fig.~\ref{6}b we also represent a partially controlled trajectory (red dots) in the phase space. This trajectory remains in the region ($-2<x_{n}<0$) as we intended. In Fig.~\ref{5}c and Fig.~\ref{6}c it is represented the corresponding controlled time series, where we also show an uncontrolled trajectory in order to compare.

\subsection{The standard map}

Here we consider the well-known standard map. This map represents the discrete dynamics corresponding to the Poincaré section of the kicked rotator system.  The system is given by:
\begin{equation}
\begin{array}{l}
y_{n}=y_{n-1}+K\sin x_{n-1}\\
x_{n}=x_{n-1}+y_{n},
\end{array}
\end{equation}
where $x_{n}$ and $y_{n}$  are taken \emph{modulo} $2\pi$.  The standard map shows hamiltonian chaos for different values of the parameter $K>0$. Depending on the initial conditions, it is possible to observe the coexistence of periodic orbits, quasiperiodic orbits, and chaotic orbits.

\begin{figure}
\includegraphics [trim=0cm 0cm 0cm 0cm, clip=true,width=1\textwidth]{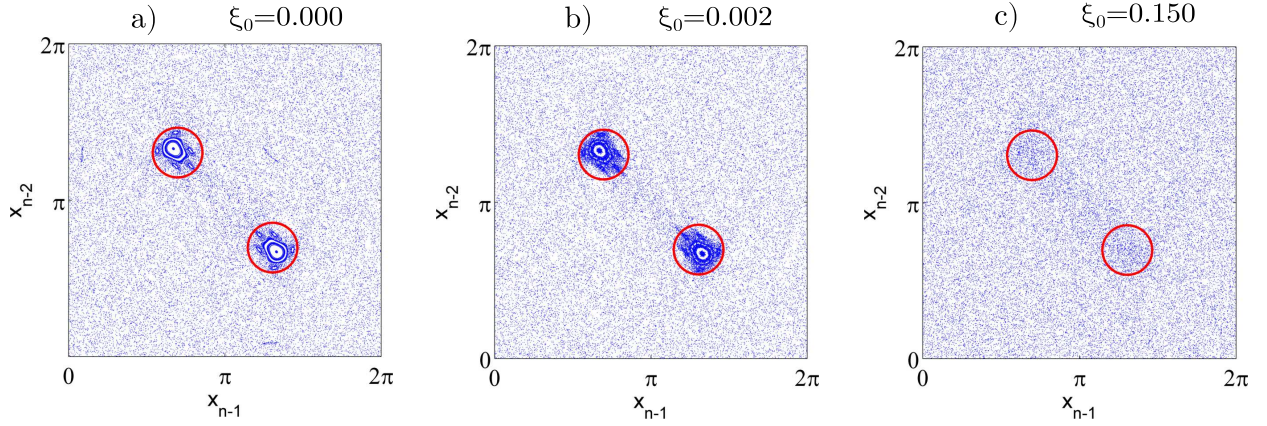}\\
\centering
\caption{\textbf{Delayed standard map ($\mathbf{x_{n}= 2x_{n-1}-x_{n-2}+K\sin x_{n-1}}$) affected by different disturbances $\xi_0$.} The points represent different trajectories in the standard map. Several initial conditions were taken to show the different dynamical behaviours (chaotic, periodic and quasiperiodic orbits). The figures represent three different cases where the trajectories are affected by random disturbances with upper bound $\xi_0=0.000$, $\xi_0=0.002$ and $\xi_0=0.150$ respectively.}
\label{7}
\end{figure}

We consider here the case $K=4.8176$. In absence of any disturbance ($\xi_0=0$), this map exhibits  chaotic regions and quasiperiodic orbits depending on the initial conditions (Fig.~\ref{7}a), however when some amount of disturbance is present ($\xi_0=0.002$), some quasiperiodic orbits vanish, and chaotic behaviour arises (Fig.~\ref{7}b).  For a large enough disturbance ($\xi_0=0.15$), no periodic or quasiperiodic orbits exist and chaotic behaviour is the only behaviour present in the system (Fig.~\ref{7}c).

  Imagine now that we want to avoid the KAM islands region in order to avoid the potential quasiperiodic behaviour of the trajectories. To do that we have applied the partial control method. In this case, the dynamics of the variable $x_n$ can be reconstructed after some arrangements, obtaining the following time delay map:
 \begin{equation}
\begin{array}{l}
x_{n}= 2x_{n-1}-x_{n-2}+K\sin x_{n-1},
\end{array}
\end{equation}
where $x_n$ is taken \emph{modulo} $2\pi$. We call this map the \emph{delayed standard map}. Considering again the disturbance and control terms, the resulting dynamics is as follows:
\begin{equation}
x_{n}=2x_{n-1}-x_{n-2}+K\sin x_{n-1} +\xi_{n}+u_n,
\end{equation}
with $|\xi_n|\leq\xi_0$ and $|u_n|\leq u_0$.

\begin{figure}
\includegraphics [trim=0cm 0cm 0cm 0cm, clip=true,width=1\textwidth]{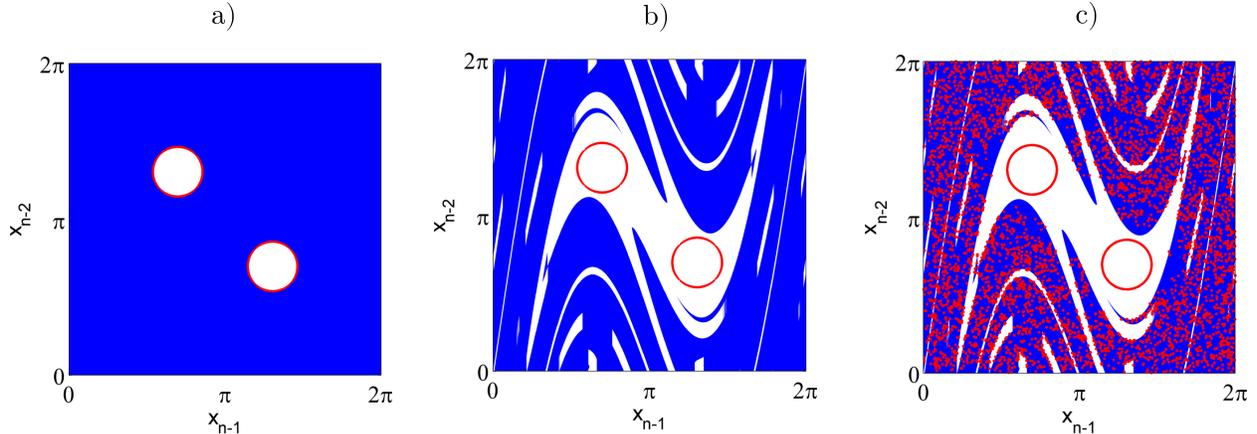}\\
\centering
\caption{\textbf{Safe set and controlled trajectory in the delayed standard map ($\mathbf{x_{n}= 2x_{n-1}-x_{n-2}+K\sin x_{n-1}}$)} a) In blue the initial region $Q_0$ that we select to keep the trajectory. b) The safe set computed with the values $\xi_0=0.15$ and $u_0=0.08$. The grid used here is $1000 \times 1000$ points. c) Partially trajectory with 5000 iterations on the safe set.}
\label{8}
\end{figure}

  In this case we consider the largest value of the disturbance ($\xi_0=0.15)$. The next step is to define the region $Q_0$ where we will keep the trajectories. This region is shown in Fig.~\ref{8}a where the two holes correspond to the KAM islands present in the deterministic case. Then, we apply the modified Sculpting Algorithm to obtain the safe set shown in Fig.~\ref{8}b. For this value of the disturbance, it was possible to control the system with controls smaller than $u_0=0.08$.  We also represent  in Fig.~\ref{8}c a partially controlled trajectory (red dots). Notice that in this case the controlled trajectory covers all the safe set due to the non dissipative dynamics of the standard map.

\subsection{The  3-dimensional hyperchaotic Hénon map}

 The partial control method in previous works, has been applied in the past to several chaotic systems, some of them very well-known like the Lorenz system or the Duffing oscillator. However, the method has never before been implemented in a hyperchaotic system which involves two or more positive Lyapunov exponents. For this reason we propose here the application of the method to the three-dimensional Hénon map \cite{HyperHenon}.

  This system  is given by:
\begin{equation}
\begin{array}{l}
x_{n}=bz_{n-1} \\
y_{n}=cx_{n-1}+bz_{n-1}\\
z_{n}=1+y_{n-1}-az_{n-1}^2.
\end{array}
\end{equation}

\begin{figure}
\includegraphics [trim=0cm 0cm 0cm 0cm, clip=true, width=0.7\textwidth]{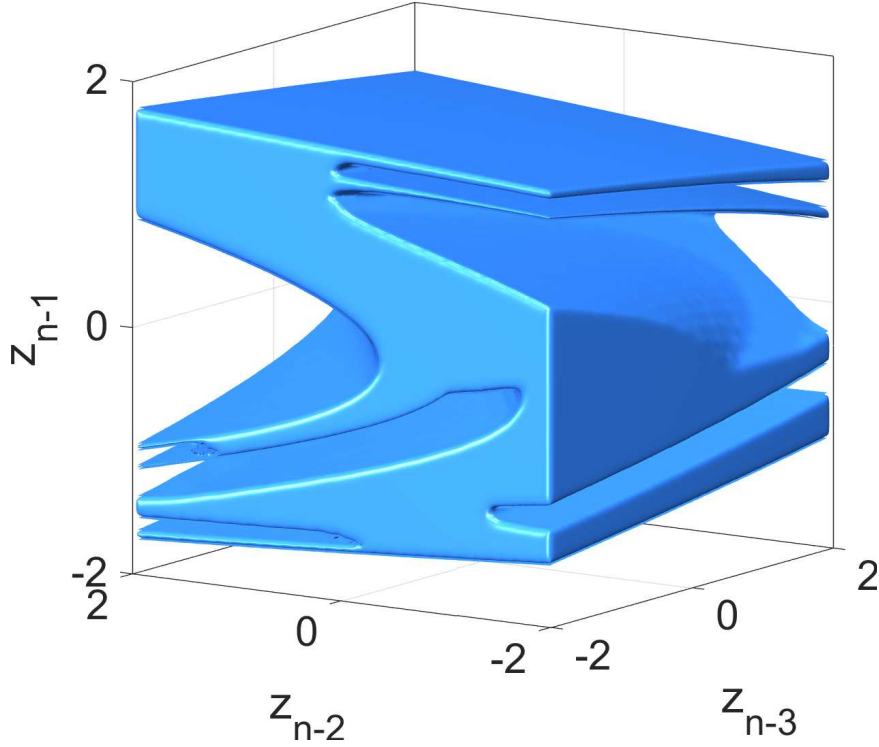}\\
\centering
\caption{\textbf{Safe set of the 3D delayed Hénon map ($\mathbf{z_{n}= 1-az_{n-1}^2+bz_{n-2}-cbz_{n-3}}$) with $\mathbf{a=1.1}$,  $\mathbf{b=0.3}$, $\mathbf{c=1}$}. A grid of $1000\times1000\times1000$ was taken in the box $[-2,2] \times [-2,2]\times [-2,2] $ that represents the initial region $Q_0$. Taking the upper bound of the disturbance $\xi_0=0.12$ and  the control $u_0=0.08$, the safe set converges after 15 iterations.}
\label{1}
\end{figure}

\begin{figure}
\includegraphics [trim=0cm 0cm 0cm 0cm, clip=true,width=0.7\textwidth]{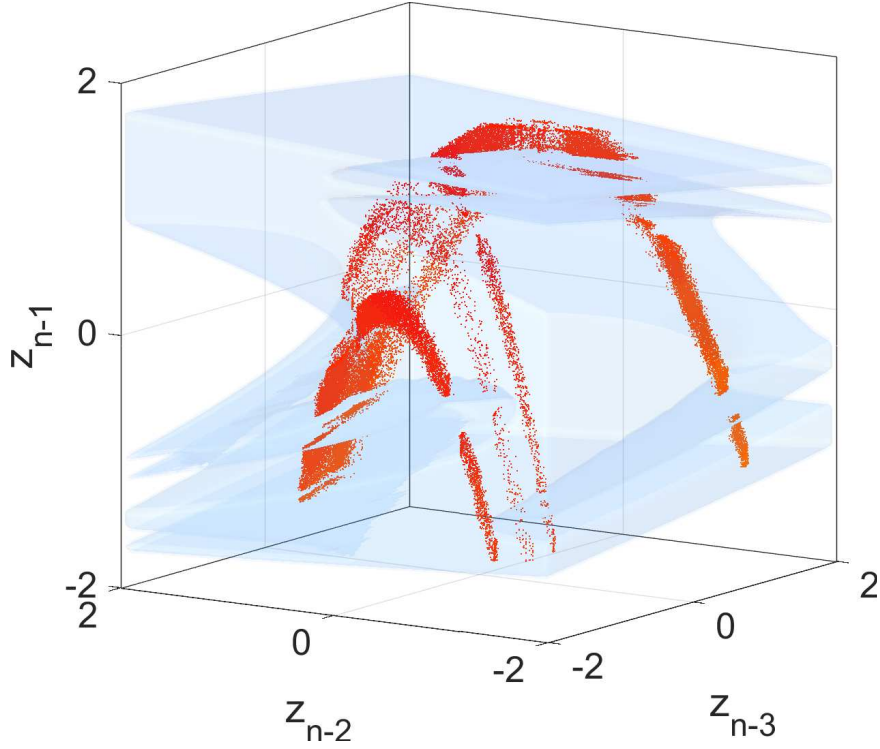}\\
\centering
\caption{\textbf{Safe set and a partially controlled trajectory in the 3D delayed Hénon map ($\mathbf{z_{n}= 1-az_{n-1}^2+bz_{n-2}-cbz_{n-3}}$) }. The safe set is represented in transparent blue to see the controlled trajectory inside (red dots).  The  variable $z_n$ is affected by a random disturbance with upper bound $\xi_0=0.12$ and  control $u_0=0.08$. }
\label{2}
\end{figure}

This map shows transient chaos for a wide range of the parameters $a$, $b$ and $c$. To compute an example, we have chosen the parameter values $a=1.1$,  $b=0.3$ and $c=1$. For these values, the trajectories with initial conditions in the box $(x_n,y_n,z_n)\in[-0.5,0.5] \times [-1,1] \times [-2,2]$ have a chaotic transient, before eventually escaping from this region towards infinity.  In this case, the effect of the disturbances in the dynamics does not change dramatically the behaviour of the trajectories. It just increases or reduces the escape time in comparison with the deterministic trajectory.

Suppose now that we have collected data from the variable $z_n$ so that we were able to reconstruct a delay-coordinate map. In this case, taking three delays is sufficient to describe correctly the dynamics of the system, that is, $z_{n}=f(z_{n-1},z_{n-2},z_{n-3})$.

The form of this delay coordinate map can be obtained by simple calculation:

\begin{equation}
\begin{array}{l}

z_{n}= 1-az_{n-1}^2+bz_{n-2}-cbz_{n-3}.

\end{array}
\end{equation}

From now on we will call this map the \emph{three-dimensional delayed Hénon map}.

In theses coordinates, values of $|z_n|>2$ involve the escape to $-\infty$  of the trajectories. In order to avoid the escape, the goal is to apply control in the variable $z_n$ to keep it in the box $(z_{n-1},z_{n-2},z_{n-3}) \in [-2,2] \times [-2,2]\times [-2,2]$.

\begin{figure}
\includegraphics [trim=0cm 0cm 0cm 0cm, clip=true,width=1\textwidth]{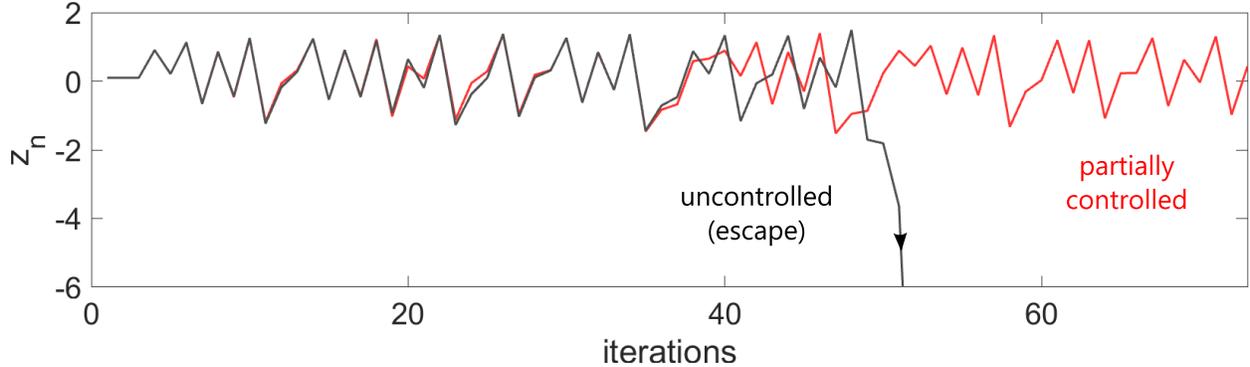}\\
\centering
\caption{\textbf{Comparison between an uncontrolled trajectory and a controlled one in the 3D delayed Hénon.} In black, the uncontrolled trajectory which after some iterations escapes to $-\infty$. In red, the controlled trajectory under the partial control scheme with $\xi_0=0.12$, $u_0=0.08$. For a fair comparison, both trajectories start with the same initial condition and are affected by the same sequence of random disturbances.}
\label{3}
\end{figure}

Introducing the disturbance term $\xi_n$ and the control term $u_n$, the partial control scheme is :
\begin{equation}
\begin{array}{l}

z_{n}= 1-az_{n-1}^2+bz_{n-2}-cbz_{n-3}+\xi_{n}+u_n,

\end{array}
\end{equation}
with $|\xi_n|\leq\xi_0$ and $|u_n|\leq u_0$. In order to show how the safe set changes depending on the disturbance value, we have computed the safe set taking $\xi_0=0.12$ and $u_0=0.08$. We have used a grid of $1000\times1000\times1000$ points covering $Q_0$, and then applied the modified Sculpting Algorithm to the safe set shown in obtained Fig.~\ref{1}.
We also represent in Fig.~\ref{2}, 10000 iterations of a partially controlled trajectory  (red dots). Notice that the trajectory remains in the box $[-2,2] \times [-2,2]\times [-2,2]$ forever. In absence of control, the trajectory abandons this box after some iterations as it is illustrated in the time series represented in Fig.~\ref{3}.

Although the variable $z_n$ was taken here as an example, in the case that the reconstructed delayed map was built with other variable $x_n$ or $y_n$, the methodology would be the same as the one presented here. The only difference would be the shape of the safe set obtained and possibly the minimum ratio $u_0/\xi_0$ achieved, since this depends on the embedded variable.

\section{Future work}

The main goal of this work is to show that partial control can also be applied in the case of delay-coordinate maps. The chaotic maps considered here are two and three-dimensional, in order to obtain two and three-dimensional delayed maps. The goal si to show clearly the control scheme proposed, but there is no restriction to apply this scheme to higher dimensional maps. We observe in general that the higher the maximal Lyapunov exponent is, the smaller the ratio $u_0/\xi_0$ necessary to control the trajectories.

The connection between time series and delay-coordinates maps obtained from an appropriate embedding, allows many different combinations regarding the constraints of the experiment. For example, the evolution of the state of the system can be taken from the measure of one variable, every $k$ iterations and the analytical expression of this evolution in terms of the delays is not always possible. Another scenario could arise if the disturbance present in the map appears in a non additive way or mixed among the variables and the parameters. In this sense, there is still much room to continue improving the control method and develop new approaches to deal with more general problems where chaos and noise are present.

\section{Conclusions}

In this work, we have shown how to apply the partial control method to different delay-coordinate nonlinear maps with chaotic behaviour and affected by random disturbances. The aim of the control scheme presented here  was to keep the chaotic trajectories in a desirable region of the phase space, applying small corrections in the observables of the system. The novelty introduced here is that, it is possible to apply the partial control method with the only knowledge and control of one variable. To achieve this goal we have modified the Sculpting Algorithm in order to find safe sets for this kind of models

 %After having built the delay-coordinate map from an appropriate embedding, the evolution of the trajectories depends on some \emph{m} delayed states and therefore the phase space associated is \emph{m}-dimensional. However, only the present state of the system can be physically controlled, so the challenge here is to control an \emph{m}-dimensional phase space perturbing only one dimension. With this constraint in mind, we have adapted the  methodology of the partial control method to implement it on this kind of systems.

 The three examples presented here, the two-dimensional cubic map, the standard map and the three-dimensional hyperchaotic Hénon map, were considered in the chaotic regime and with some disturbances affecting them.  By applying a smaller control $u_0<\xi_0$, we have shown that it is possible to keep the trajectories within a desirable region of the phase space. In this sense we want to recall that the desirable regions selected here to maintain the trajectories were only examples, and many other choices are possible depending on our control convenience. We have also applied for the first time the partial control method to a hyperchaotic map. Another interesting result of this work is the dynamics of a partially controlled Hamiltonian system. As we have shown it covers the whole safe set. This is a fundamental difference with the dissipative case.

Finally, although we consider here mathematical models to express the maps, we believe that the method can be applied in the same way to delay-coordinate maps built from experimental time series. That would be the next step in the development of this control method.

\begin{acknowledgments}
This work was supported by the Spanish Ministry of Economy and Competitiveness
under Project No. FIS2013-40653-P and by the Spanish State Research Agency (AEI)
and the European Regional Development Fund (FEDER) under
Project No. FIS2016-76883-P. MAFS acknowledges the
jointly sponsored financial support by the Fulbright Program
and the Spanish Ministry of Education (Program No. FMECD-ST-2016).

\end{acknowledgments}

\end{document}